\documentclass[9pt,a4paper,twocolumn]{article}
\usepackage[dvips]{graphicx}

\usepackage{amssymb,amsmath}
\usepackage{fancyheadings}
\newcommand{\mib}[1]{\mbox{\boldmath$#1$}}
\newcommand{\tx}[1]{\textrm{\scriptsize #1}}

\usepackage{xcolor}
\newcommand{\rev}[1]{\textcolor{black}{#1}}

\title{Dynamic mode decomposition to retrieve torsional Alfv\'{e}n waves}

\author{K. Hori, Graduate School of System Informatics, Kobe University,
 Kobe 657-8501, Japan.\\
S.~M. Tobias, Department of Applied Mathematics, University of Leeds,
 Leeds  LS2 9JT, UK.\\
R.~J. Teed, School of Mathematics and Statistics, University of Glasgow,
 Glasgow G12 8SQ, UK.}

\date{}

\begin{document}
\maketitle

\begin{minipage}{.43\textwidth}
{\bf Abstract} 
Dynamic mode decomposition (DMD) is utilised to identify the intrinsic
 signals arising from planetary interiors. 
Focusing on an axisymmetric quasi-geostrophic magnetohydrodynamic (MHD) wave
 -- called torsional Alfv\'{e}n waves (TW) -- we examine the utility of DMD
 in two types of MHD direct numerical simulations:
 Boussinesq magnetoconvection 
 \rev{and anelastic convection-driven dynamos}
 in rapidly rotating spherical shells\rev{, which model the dynamics in Earth's core and in Jupiter, respectively.}
We demonstrate that
DMD is capable of distinguishing internal modes and boundary/interface-related modes
 from the timeseries of the internal velocity.
Those internal modes may be realised as free TW, in terms of eigenvalues and eigenfunctions of their normal mode solutions.
Meanwhile it turns out that, in order to account for the details,
 the global TW eigenvalue problems in spherical shells need to be further addressed.


\end{minipage}

\vspace{1em}
\section{Introduction}

Signals arising from the deep interiors of planets
 are crucial sources of information about their physical conditions, information that is essentially inaccessible \textit{in situ}.
For instance, 
 revealing the core dynamics at the centre of Earth is very limited:
 it is in principle inferred from the magnetic activity observed at/above the surface
 and there is difficulty in extracting details of its variations originated in the fluid core. 
Whilst satellite measurements have advanced over the last decades to find its "rapid" dynamics of \rev{interannual changes}
 the data investigation still largely relies on \rev{ground-based records \rev{over decades and longer} (a recent review by \cite{G19} and references therein)}.Now the archaeo- or palaeo- magnetic datasets covered a wider area of the globe for longer intervals \cite{NHKSH14}; 
 nonetheless their inhomogeneous coverage is inevitable. 

An efficient technique to extract any intrinsic signals from the limited timeseries is therefore necessary to reveal the dynamics in the planetary interior and hence the dynamo.
In addition to the classic Fourier transformation,
 earlier studies attempted the wavelet transformation \cite{AGHLS95} and the empirical mode decomposition (EMD) \cite{RYR07,SJM12} comprising cubic splines.
Data-driven, modal analyses by the proper orthogonal decomposition (POD) --also known as principle component analysis (PCA) or empirical orthogonal functions (EOF)-- have been made over geomagnetic datasets \cite{CM14,DPJM19}.
POD was also used for model reduction of direct stochastic simulations for rotating fluids \cite{ATM18}. 
The method is however based on the energy ranking, with which any tiny, but physically relevant, signals may be lost.
In many cases a POD mode comprises several modes which have different frequencies and growth rates.

Dynamic mode decomposition (DMD)\cite{S10} -- an algorithm to approximate modes of the Koopman operator \cite{RMBSH09}-- enables an extended, equation-free analysis to represent the temporal structures superior to POD,
 for example, by separating into individual mode of frequency and growth rate. 
The technique is now providing a powerful tool for diagnostics, future prediction, and control of
 multi-dimensional systems in a broad range of application areas, in addition to  fluid dynamics (see \cite{KBBP16} and references therein).

As observational exploration has developed,
 there is increasing evidence of magnetohydrodynamic (MHD) waves
 excited in Earth's fluid core.
A special class that is axisymmetric and quasi-geostrophic
 ($\mib{\Omega}\cdot \mib{k}\approx 0$ with $\mib{\Omega}$ being the rotation axis and $\mib{k}$ being the wavenumber vector)
 is called torsional Alfv\'{e}n waves (TW):
 the inviscid, ideal wave for a Boussinesq/anelastic fluid is governed by  
 \begin{equation}
   \frac{\partial^2}{\partial t^2} \frac{\langle u'_\phi \rangle}{s}
  = \frac{1}{s^3 h \langle \rho \rangle}
       \frac{\partial}{\partial s} \left( s^3 h \langle \rho \rangle U_\tx{A}^2
       \frac{\partial}{\partial s} \frac{\langle u'_\phi \rangle}{s}  \right)  \label{eq:TW}
 \end{equation}
\cite{Bra70,RA12,TJT15,HTJ19} where $s$ is the cylindrical radius, $h$ is the height from the equator,
 $\langle u'_\phi \rangle$ is the axisymmetric geostrophic component of the fluctuating azimuthal velocity,
 $\langle \rho \rangle$ is the axisymmetric geostrophic component of the background density, 
 and $U_\tx{A} = \langle B_s \rangle/\sqrt{\mu\langle \rho \rangle }$ is the Alfv\'{e}n speed 
 with $\langle B_s \rangle$ being the equivalent component of the radial magnetic field and $\mu$ being the permeability. 
Although the excitation of these waves in the fluid core has been long a subject of debate 
the waves are believed to account for a several-year signal 
 that was found in core flow models inverted from the observed geomagnetic secular variation \cite{GJCF10,G19}. 
 The signal was not very strong so that the identification required bandpass filterings.
The equivalent periodicity in geomagnetic time-sequences was exploited by EMD and the Fourier analysis \cite{SJM12}.

This MHD wave could reasonably be excited in other rapidly-rotating planets,
 such as in Jupiter's metallic hydrogen region \cite{HTJ19}. 
Here, unlike for a terrestrial planet, the disturbances may penetrate through the gaseous envelope to be visible near the surface. 
The cloud deck has been monitored by ground-based campaigns to find its long-term variation\rev{.
S}everal-year periodicities in brightness were visualised 
 by Lomb-Scargle and wavelet analyses at each latitude
 and were addressed by POD over the whole latitudes \cite{AFetal19}.

We here examine how DMD could help to identify signals of TW
 by means of spherical DNS for MHD fluids.
For a first attempt we concentrate on a rather small set of data and on standard DMD.

\vspace{1em}
\section{A brief overview of DMD}

We consider standard DMD \cite{S10}, incorporating some recent updates \cite{JSN14,HS17}, in this study.
We consider a data sequence $[ \mib{\psi}_0, \mib{\psi}_1, ..., \mib{\psi}_N]$
 where $\mib{\psi}_i = \psi (x, i\Delta t)$ is a real vector with $N+1$ points in time $t$ sampled every $\Delta t$
 and $M$ components in space $x$, $\psi_i \in \mathbb{R}^M$, 
 with $M \rev{>} N$. 
We then form two matrices:
\begin{equation}
 \begin{split}
  \Psi_0 &= [ \mib{\psi}_0 \; \mib{\psi}_1 \; ... \; \mib{\psi}_{N-1}] \in \mathbb{R}^{M\times N}  \\
 \textrm{and} \quad  
  \Psi_1 &= [ \mib{\psi}_1 \; \mib{\psi}_2 \; ... \; \mib{\psi}_{N}  ] \quad \in \mathbb{R}^{M\times N}, 
  \end{split}
\end{equation}
and link the two so that the neighbouring snapshots can be expressed
 as a linear, discrete-time system:
\begin{equation}
 \Psi_1 = A \Psi_0  \quad \textrm{where} \quad   A \in \mathbb{R}^{M\times M}  \; .    \label{eq:originalA}
\end{equation}
Given the $M$ eigensolutions of $A =\Psi_1 \Psi_0^\dag$ (where $^\dag$ denotes the pseudo inverse matrix),
 the data sequence may be represented by the linear combination of those modes 
 (sometimes called exact DMD modes). 

For a quite large set of data, which is the case of interest, 
 we may seek an optimal representation $\widetilde{A} \in \mathbb{R}^{r\times r}$ of the matrix
 (with $r < M$) such that
\begin{equation}
 A \approx U \widetilde{A} U^\tx{T}   \qquad \textrm{or} \qquad U^\tx{T} A U \approx \widetilde{A}
  \label{eq:tildeA}
\end{equation}
where $U$ is obtained from an 
singular value decomposition of $\Psi_0$, i.e. 
\begin{equation}
 \Psi_0 = U \Sigma V^\tx{T}   \; .
\end{equation}
Here $r$ is the rank of $\Psi_0$,  
 $\Sigma$ is an $r\times r$ diagonal matrix with diagonal entries $\sigma_j \ge 0$ ordering as 
 $\sigma_1 \ge \sigma_2 \ge  ... \ge \sigma_r$,
\begin{equation}
 \begin{split}
  U &= [ \mib{u}_1  \; \mib{u}_2 \; ... \mib{u}_r ] \in \mathbb{R}^{M \times r} , \\ 
 \textrm{and} \quad  
  V &= [ \mib{v}_1  \; \mib{v}_2 \; ... \mib{v}_r ] \in \mathbb{R}^{N \times r} . \end{split}
\end{equation}
Note $U$ contains the spatial structures (topos) whilst $V$ contains the temporal structures (chronos). 
Vectors $\mib{u}_j$ in $U$ may be regarded as POD modes of the dataset in $\Psi_0$,
 together with the singular values $\sigma_j$ describing the contained energy in descending order. 
The matrix $\widetilde{A}$ is obtained by minimising the Frobenius norm of the difference 
\begin{equation}
 || \Psi_1 - A \Psi_0 ||^2_\tx{F}
  \approx || \Psi_1 - U \widetilde{A} \Sigma V^\tx{T} ||^2_\tx{F}
		\label{eq:residual}
\end{equation}
to give 
\begin{equation}
 \widetilde{A} = U^\tx{T} \Psi_1 V \Sigma^{-1} .
\end{equation}
When the $r$ eigenvalues ($\mu_j$) and eigenvectors ($\mib{z}_j$) 
 of $\widetilde{A}$ are found,
 the original data at time $i\Delta t$ may be approximated as   
\begin{equation}
 \mib{\psi}_i  \approx \sum_{j=1}^{r} U \mib{y}_j \; \mu_j^i \;  \mib{z}_j^\tx{\rev{H}} \mib{x}_0
               \equiv   
                       \sum_{j=1}^r  \mib{\phi}_j  e^{ \lambda_j t} \alpha_j  \; .
\end{equation}
Here the (projected) DMD eigenvalue is given by $\lambda_j = (\ln{\mu_j})/\Delta t$,
 its eigenfunction $\mib{\phi}_j = U \mib{y}_j$ with $\mib{y}_j$ satisfying $\mib{z}^\tx{\rev{H}}_k \mib{y}_j = \delta_{jk}$ \rev{(where the superscript $^\tx{H}$ denotes the complex conjugate transpose)}, and 
 its optimal amplitude $\alpha_j = \mib{z}_j^\tx{\rev{H}} \mib{x}_0$ with
 $\mib{x}_0$ being the initial data \rev{where $\mib{x}_{i+1} = \widetilde{A} \mib{x}_i$}. 
The real and imaginary parts of $\lambda_j$ represent the growth rate and the frequency, respectively.
%

%
The optimal amplitude is determined so that the solutions should minimise
 \begin{equation}
  || \Psi_0 - \Phi D_\alpha V_\tx{and} ||^2_\tx{F}    \label{eq:res1}
 \end{equation}
 where $\Phi = [\mib{\phi}_1 \; \mib{\phi}_2 \; ... \mib{\phi}_r ] \in \rev{\mathbb{C}}^{M\times r}$,
 $D_\alpha$ is an $r\times r$ diagonal matrix with diagonal entries $\alpha_j$,
 and $\{ V_\tx{and} \}_{ji} = \mu_j^i \in \rev{\mathbb{C}}^{r\times N}$.
This may be rewritten as minimising
\begin{equation}
 || \Sigma V^\tx{T} - Y D_\alpha V_\tx{and} ||^2_\tx{F} \; ,  \label{eq:res2}
\end{equation}
where $\Phi = U Y$,  or minimising
\begin{equation}
 J (\mib{\alpha}) = \mib{\alpha}^\tx{\rev{H}} P \mib{\alpha} - q^\tx{\rev{H}} \mib{\alpha} - \mib{\alpha}^\tx{\rev{H}} q + s 
\end{equation}
where $\mib{\alpha} = [\alpha_1 \; \alpha_2 \; ... \; \alpha_r ]$,
 $P = (Y^\tx{\rev{H}} Y)\circ (\overline{ V_\tx{and} V_\tx{and}^\tx{\rev{H}} })$,
 $q = \overline{\textrm{diag} (V_\tx{and} V \Sigma^\tx{T} Y) }$,
 and $s = \textrm{trace}(\Sigma^\tx{T}\Sigma)$. 
Here the overbar denotes the complex conjugate, and $\circ$ represents the element-wise multiplication. 
By minimising $J(\mib{\alpha})$
 the optimal vector of the DMD amplitude is obtained as 
\begin{equation}
 \mib{\alpha} = P^{-1} q \; .
\end{equation}

In the analysis presented below, we quantify the accuracy of our decomposition
 in terms of the residual, (\ref{eq:res1}) or (\ref{eq:res2}), between the sampled and estimated data
 and also the performance of loss  
\begin{equation}
 \Pi_\tx{loss}  \equiv 100 \sqrt{\frac{J(\mib{\alpha})}{J(\mib{0})}}   \label{eq:Ploss}
\end{equation}
of the retained DMD modes. 
The efficiency of DMD can also be measured by the rank $r$ of the approximated matrix
 against  $M$ for the sampled data. 
In addition, we check the singular values $\sigma_j$ 
 to learn how many DMD (or POD) modes, out of the $r$ modes, will effectively approximate the data.

\vspace{1em}
\section{DMD of numerical data}

We consider \rev{two} types of numerical data which were obtained
 from 3D DNS of MHD \rev{fluids} 
 in rapidly rotating spherical shells: 
 (i) Boussinesq convection permeated by a dipolar poloidal field (referred to as 'magnetoconvection' hereafter) \cite{TJT15}
 and (ii) dynamos driven by anelastic convection incorporating a transition to the hydrodynamic envelope ('Jovian dynamo') \cite{J14,HTJ19}.
\rev{The respective models were designed for Earth's liquid iron core and Jupiter's metallic/molecular hydrogen region, respectively.} 
We note that the dynamo runs presented below yield self-generated magnetic fields dominated by an axial dipole. For both types of models,
time and length are scaled, respectively, by magnetic diffusion time $\tau_\tx{m}$
 and by the radius $r_\tx{o}$ of the outer shell in our simulations. 
This radius is identified the top of Earth's core for the magnetoconvection cases, 
 and about 0.96 of Jupiter's nominal radius for the Jovian dynamo cases.

\begin{figure}[h]
\begin{center}
 \includegraphics[width=0.7\linewidth]{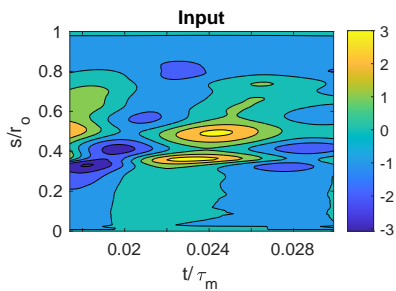}
\end{center}
\vspace{-2em}
 \caption{Axisymmetric, geostrophic zonal flow $\langle u_\phi \rangle$ in a magnetoconvection case 5E6P.1d \cite{TJT15}. In the northern hemisphere.} 
 \label{fig:data}
\end{figure}

The original spherical data were converted to cylindrical coordinates $(s,\phi,z)$
 and averaged over $\phi$ and $z$
 to focus on the axisymmetric, geostrophic component $\langle u_\phi \rangle (s\rev{,t})$.
In the all runs analysed below,
 the cylindrical fluctuations
 were previously identified as travelling \rev{or standing} in $s$ and compared with the predicted phase/ray paths of TW \cite{TJT15,HTJ19}:
 this may be regarded a local approach. 
An example for magnetoconvection is exhibited in figure \ref{fig:data}.
The contours of $\langle u_\phi \rangle$ in $s$-$t$ space show
 they mostly fluctuate outside the tangent cylinder (TC), located at $s/r_\tx{o} = 0.35$, 
 and travel outwards.
Below we shall examine if such TW identification in $\langle u_\phi \rangle$
 is endorsed by DMD and TW normal modes:
 we call this a global approach, contrasting with the local one.

\vspace{1em}

\vspace{1em}
\begin{figure}[h]
\begin{center}
 \includegraphics[bb= 42mm 2mm 92mm 65mm, clip, width=0.35\linewidth]{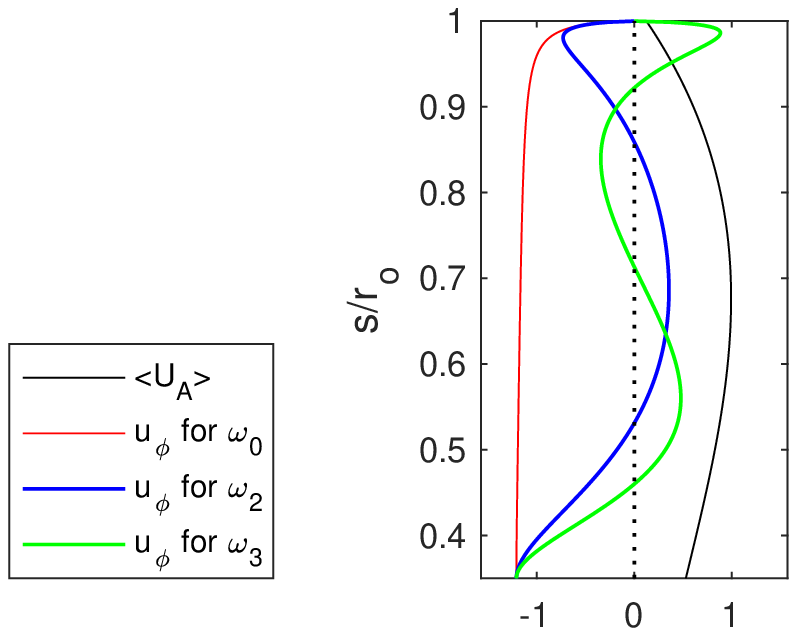}
\end{center}
\vspace{-2em}
 \caption{Profiles of the normalised Alfv\'{e}n speed $U_\tx{A}$ in run 5E6P.1d (black solid curve) and 
 the eigenfunctions $\langle u'_\phi \rangle$ of the 0th (red), 2nd (blue), and 3rd (green) TW normal modes \rev{outside the TC}, provided a normalising factor at the inner bound.}
 \label{fig:TW_normal_modes}
\end{figure}

\vspace{1em}
\begin{figure}[h]
\begin{center}
 \includegraphics[width=0.7\linewidth]{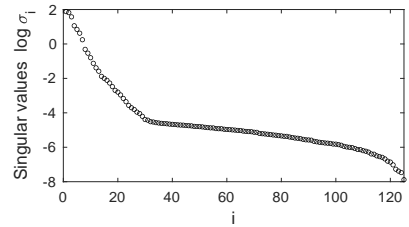}
\end{center}
\vspace{-2em}
 \caption{\rev{Singular values $\sigma_j$ in run 5E6P.1d.} } 
 \label{fig:singular_values}
\end{figure}

\vspace{-1em}
\subsection{Magnetoconvection}

For a given Alfv\'{e}n speed $U_\tx{A}(s)$ 
 and considering normal mode solutions ($\propto \exp{\textrm{i} \omega t}$),
 the TW equation (\ref{eq:TW}) is solved outside the TC
 with the Matlab routine bvp4c. 
The number of grid points in $s$ is 500.
The numerical code is benchmarked with solutions in a whole sphere \cite{RA12,MJ16}:
 the first four eigenvalues for constant $U_\tx{A}$ agree with those in the previous work to five decimal places.
The first six eigenvalues (denoted by $\omega_i$) for the current problem,
 $U_\tx{A}(s)$ obtained in the DNS, 
 are found to be 77.2, 479, 731, 1000, 1275, and 1553,
 provided free-slip and no-slip conditions at $s/r_\tx{o} =$ 0.3501 and 0.9999, respectively.
Figure~\ref{fig:TW_normal_modes} demonstrates profiles of the assumed background profile $U_\tx{A}(s)$
 and selected eigenfunctions corresponding to the calculated eigenvalues.
Here the $i$-th normal mode has $i$ nodes in $s$.

These TW normal modes are compared with DMD modes of the data
 $\langle u_\phi \rangle (s,t)$.
The dataset has 127 grid points \rev{($=M$) in $s$. The time sequence is 0.03 long with the sampling time $\Delta t = 0.0001$,
 totaling up to 301 snapshots. 
Varying the number of analysed snapshots,
 we confirm 125 snapshots ($=N$) spanning over the window $0.0125 \le t/\tau_\tx{m} < 0.03$
 (fig.~\ref{fig:data}), 
 gives the rank $r = 125$ 
 and yield the residual (\ref{eq:res2}) no greater than $10^{-6}$. 
Then} $\Pi_\tx{loss}$ (\ref{eq:Ploss}) is diminished to less than 1\% when 13 modes,
 out of the 125 DMD modes, are retained. 
The singular values $\sigma_j$ \rev{against $j$ are plotted in Figure~\ref{fig:singular_values} to}
 show a sharp change at $j \approx 30$,
 beyond which the slope becomes gentle:
 this implies the low-dimensional structures are present. 
We also \rev{verify} the projected DMD modes indeed pose a subset of the entire $M$ eigenmodes of $A$ (\ref{eq:originalA}),
 which are explicitly computed from the pseudo inverse matrix $\Psi_0^\dag$.
The eigenvalues, $\mu_j$, almost shape a circle in the complex plane (not shown).

\vspace{1em}
\begin{figure}[h]
\begin{center}
 \includegraphics[bb= 10mm 4mm 110mm 139mm, clip, width=0.7\linewidth]{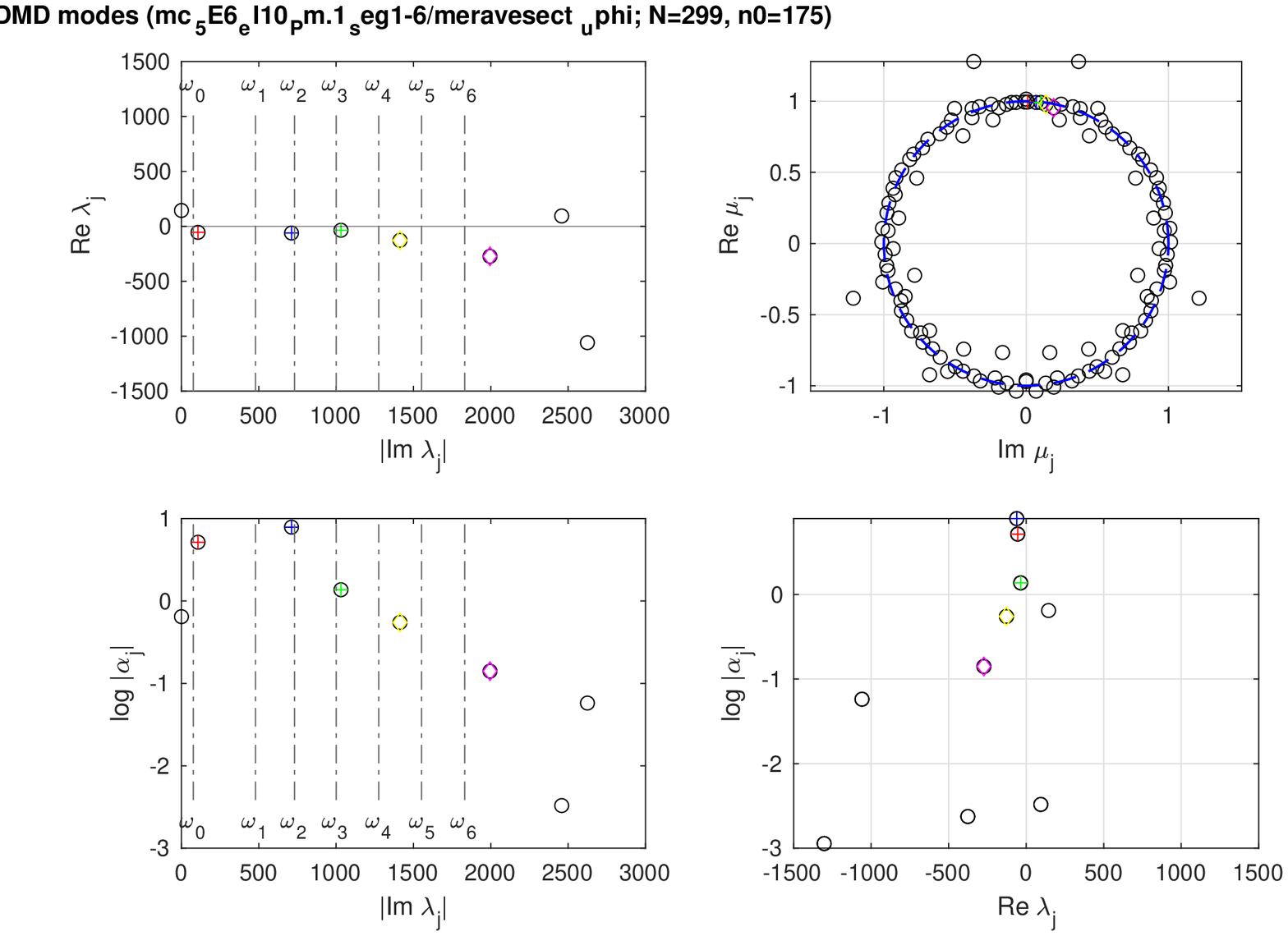} 
\end{center}
\vspace{-2em}
 \caption{DMD eigenvalues (top) and amplitude (bottom) in run 5E6P.1d. Modes 1, 2, 3, 4, and 5 are
 highlighted in red, blue, green, yellow, and magenta, respectively. 
 The \rev{vertical} dashed-dotted line labelled by $\omega_i$ indicates the frequency of the $i$-th TW normal mode.}
 \label{fig:eigenvalues}
\end{figure}

\vspace{1em}
\begin{figure}[h]
\begin{center}
 \includegraphics[bb= 15mm 12mm 100mm 192mm, clip, width=0.58\linewidth]{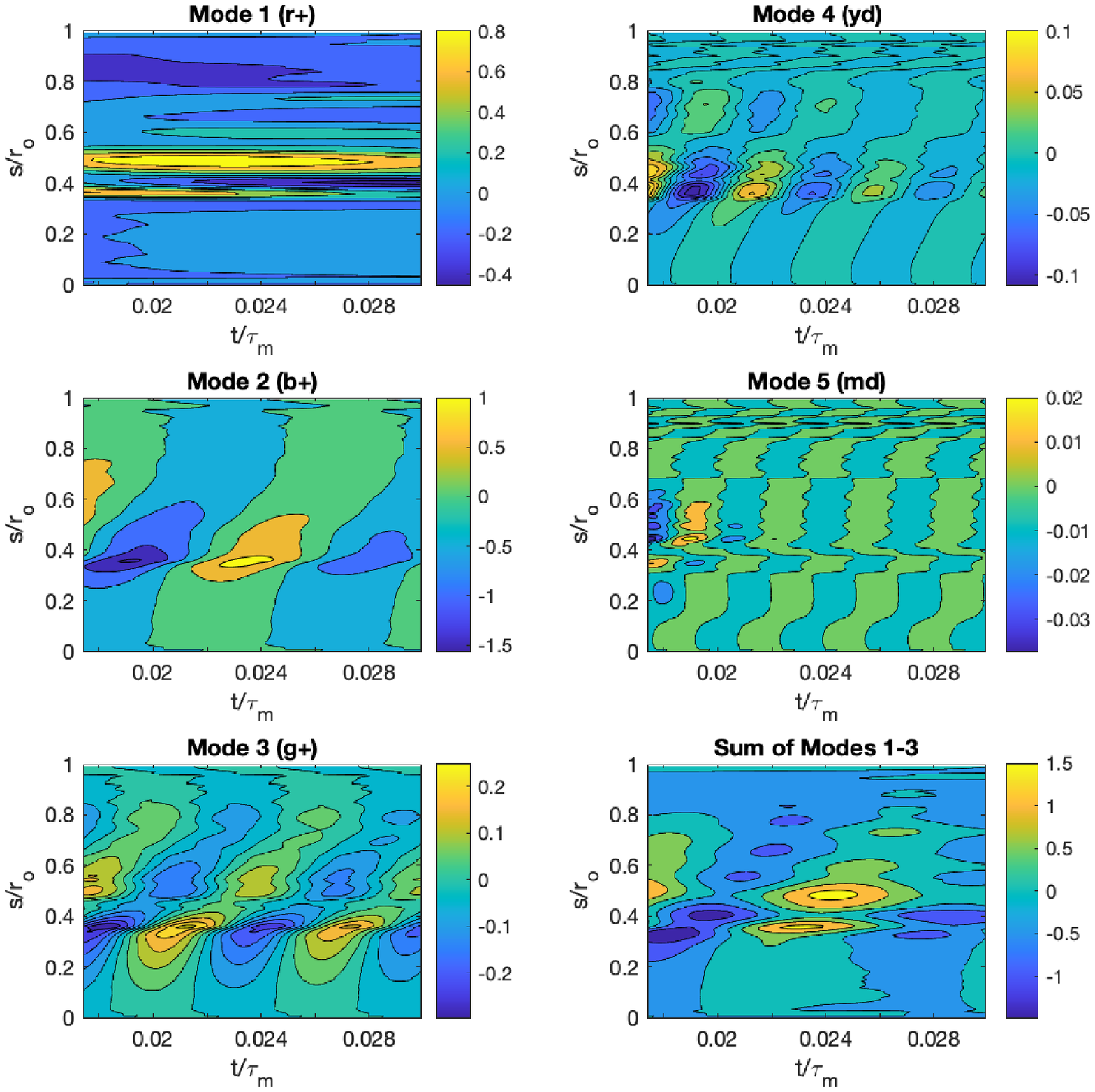}
\end{center}
\vspace{-2em}
 \caption{Spatiotemporal structures of DMD Modes 1-3 (from top to bottom) in run 5E6P.1d.}
 \label{fig:eigenfunctions_s-t}
\end{figure}

\vspace{1em}
\begin{figure}[h]
\begin{center}
 \includegraphics[bb= 28mm 3mm 78mm 72mm, clip, width=0.3\linewidth]{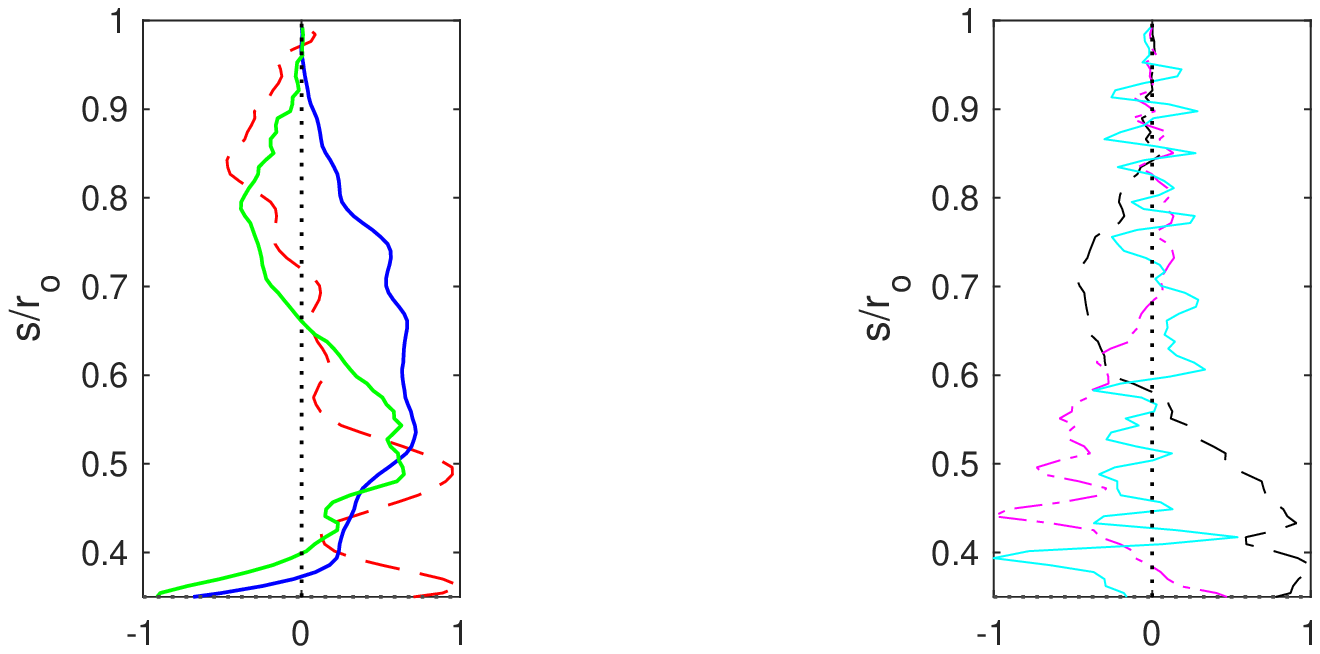}
\end{center}
\vspace{-2em}
 \caption{Profiles of DMD eigenfunctions of Modes 1 (red \rev{dashed} curve), 2 (blue \rev{solid}), and 3 (green \rev{solid}) in run 5E6P.1d.
 Only outside the TC is shown.}
 \label{fig:eigenfunctions}
\end{figure}

The top panel of figure~\ref{fig:eigenvalues} displays the DMD eigenvalues, $\lambda_j = (\ln{\mu_j})/\Delta t$,
 of the data in frequency-growth rate space, 
 whilst the amplitude $|\alpha_j|$ as a function of frequency is shown in the bottom panel.
In order to seek a stable wave solution,
 we focus on modes that have nonzero frequencies in the range of the expected TW normal modes
 and small growth rates (or their quality factors, $Q=|\textrm{Im}(\lambda_j)/2\textrm{Re}(\lambda_j)|$,
 are no lower than unity). 
In both panels
 the frequencies $\omega_i$ of the first six TW modes are indicated by vertical dashed-dotted lines. 
The DMD extracts a non-oscillating component (zero frequency but growth rate $\approx 144$)
 and visualises several significant modes (which are highlighted in colour). 
The slowest mode (red; referred to as `Mode 1') has a frequency close to $\omega_0$;
 the strongest is the second slowest mode (blue; `Mode 2')
 of frequency $\approx 711.6$, close to $\omega_2$, and growth rate $\approx -60.9$.  
The third slowest mode (green; `Mode 3') of frequency $\approx 1031$ and growth rate $\approx -35.5$
 is close to $\omega_3$.
The fourth mode (yellow; `Mode 4') is found in between $\omega_4$ and $\omega_5$;
 there are more modes at higher frequency, including Mode 5 (magenta) beyond $\omega_6$.

We further exploit eigenfunctions of those DMD modes
 to distinguish the free oscillation.
Figure~\ref{fig:eigenfunctions_s-t} demonstrates contours of
 the selected DMD eigenfunctions, $\phi_j \exp{(\lambda_j t)} \alpha_j$, in $s$-$t$ space.
Mode 1 appears to be quasi-steady and relevant in the neighbourhood of the TC.
By contrast, Modes 2 and 3 exhibit a travelling nature towards the equator $s/r_\tx{o} = 1$.
Their radial structures, $\phi_j$, are focused in figure~\ref{fig:eigenfunctions}.
The red dashed curve confirms Mode 1 representing a TC-related mode.  The
other two curves show local maxima within the internal region: 
 Mode 2 (blue solid curve) has a crest at $s/r_\tx{o} \approx 0.55$ and crosses zeros at $s/r_\tx{o} \approx 0.4$; 
 Mode 3 (green) adds another trough at $s/r_\tx{o}\approx 0.8$ and another crossing at $s/r_\tx{o}\approx 0.65$.
We therefore regard Modes 2 and 3 as related to an internal, free oscillation. 
Superposition of Modes 2 and 3 may represent the essential part of the wave motion (not shown). 
\rev{The two modes are retrieved even when the the number $N$ of analysed snapshots is limited to 75
 (the window $0.0224 \le t/\tau_\tx{m} < 0.03$),
 whilst Mode 2 is only found to be a quasi-stationary wave with 25 snapshots ($0.0274 \le t/\tau_\tx{m} < 0.03$).}

We however note that the number of the nodes does not completely match
 that expected from the 2nd or 3rd TW normal mode (fig.~\ref{fig:TW_normal_modes})
 despite the agreement in the frequency (fig.~\ref{fig:eigenvalues} top).
Indeed, as $s/r_\tx{o}  \rightarrow 1$,
 the discrepancy in eigenfunction between the DMD and the TW normal modes becomes evident\rev{,
 indicating a damping occurs in the DNS.}
These could arise from the oversimplified model adopted for computing the TW normal modes: equation
 (\ref{eq:TW}) clearly has a singularity at the outer bound. 
The inclusion of resistivity and/or viscosity may avoid the issue
 and reasonably influence details of the normal modes \cite{MB07}.\\

\vspace{1em}
\begin{figure}[h]
\begin{center}
 \includegraphics[bb= 37mm 3mm 88mm 76mm, clip, width=0.3\linewidth]{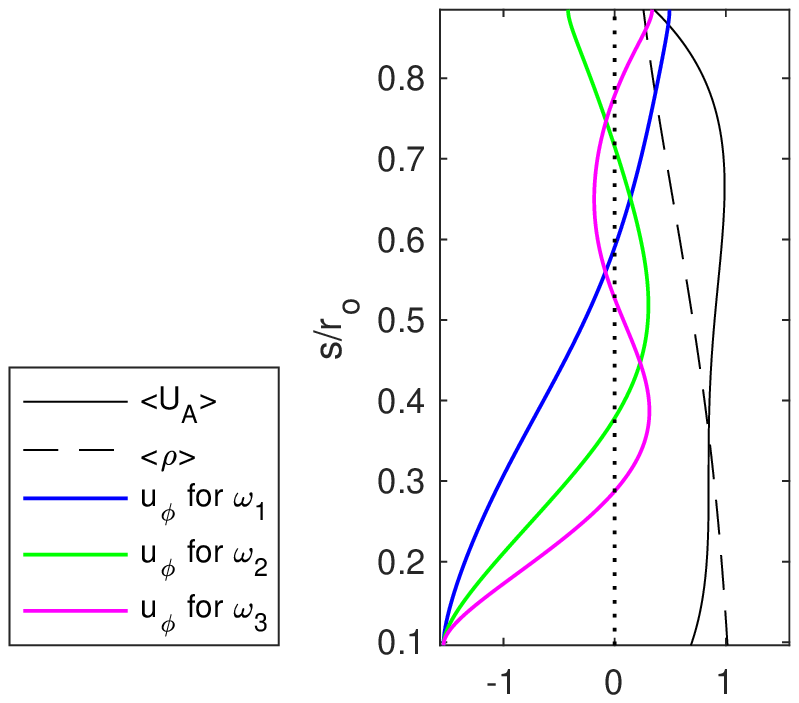}
\end{center}
\vspace{-2em}
 \caption{\rev{Profiles of the normalised Alfv\'{e}n speed $U_\tx{A}$ (black solid curve)
 and density $\langle \rho \rangle$ (black dashed) in \rev{a Jovian dynamo run E \cite{J14,HTJ19}}, and 
 the eigenfunctions $\langle u'_\phi \rangle$ of the 1st (blue), 2nd (green), and 3rd (magenta) TW normal modes \rev{within the metallic region},
 provided a normalising factor at the inner bound.}}
 \label{fig:TW_normal_modes_jup}
\end{figure}

\vspace{1em}
\begin{figure}[h]
\begin{center}
 \includegraphics[width=0.7\linewidth]{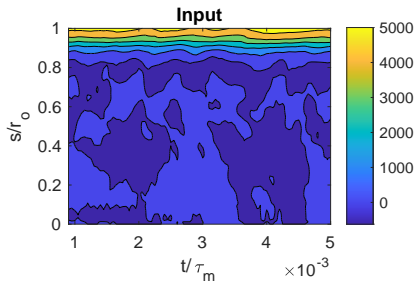}
\end{center}
\vspace{-2em}
 \caption{Axisymmetric, geostrophic zonal flow $\langle u_\phi \rangle$ in \rev{run E}. In the northern hemisphere. The transition from the metallic to molecular region begins at \rev{about $s/r_\tx{o} = 0.885$-$0.94$}.}
 \label{fig:data_jup}
\end{figure}

\subsection{Jovian dynamo}

For the Jovian dynamo cases,  a more complex structure of the background profiles is imposed \rev{(see details in \cite{J14})}.
The density and electrical conductivity vary with radius
 and drastically\rev{, but smoothly,} decrease across a transition zone from the metallic hydrogen to the molecular hydrogen regions:
 the transition 
 begins at about \rev{$s/r_\tx{o} = 0.885$-$0.944$}.
\rev{The TC attached to a rocky core is set to be at $s/r_\tx{o} = 0.0963$.}
When the simple TW model (\ref{eq:TW}) is applied to the metallic region only ($0.0963 \le s/r_\tx{o} \le 0.885$)
 the normal modes for $U_\tx{A}$ in the run 
 and free slip boundary conditions have frequencies $\omega_i$,
 from the lowest, of 4804.8, 7716.6, 10492, 13308, 16179, and 19094.
Their eigenfunctions are amplified with decreasing $s$, or the density
 \rev{(see selected ones in figure~\ref{fig:TW_normal_modes_jup})}. 
Those precalculations would guide us to interpret DMD signals shown below;  
 we however caution the current model omits any coupling between the metallic region and the overlying transition \rev{layer}.

\begin{figure}[h]
\begin{center}
 \includegraphics[bb= 10mm 5mm 110mm 140mm, clip, width=0.7\linewidth]{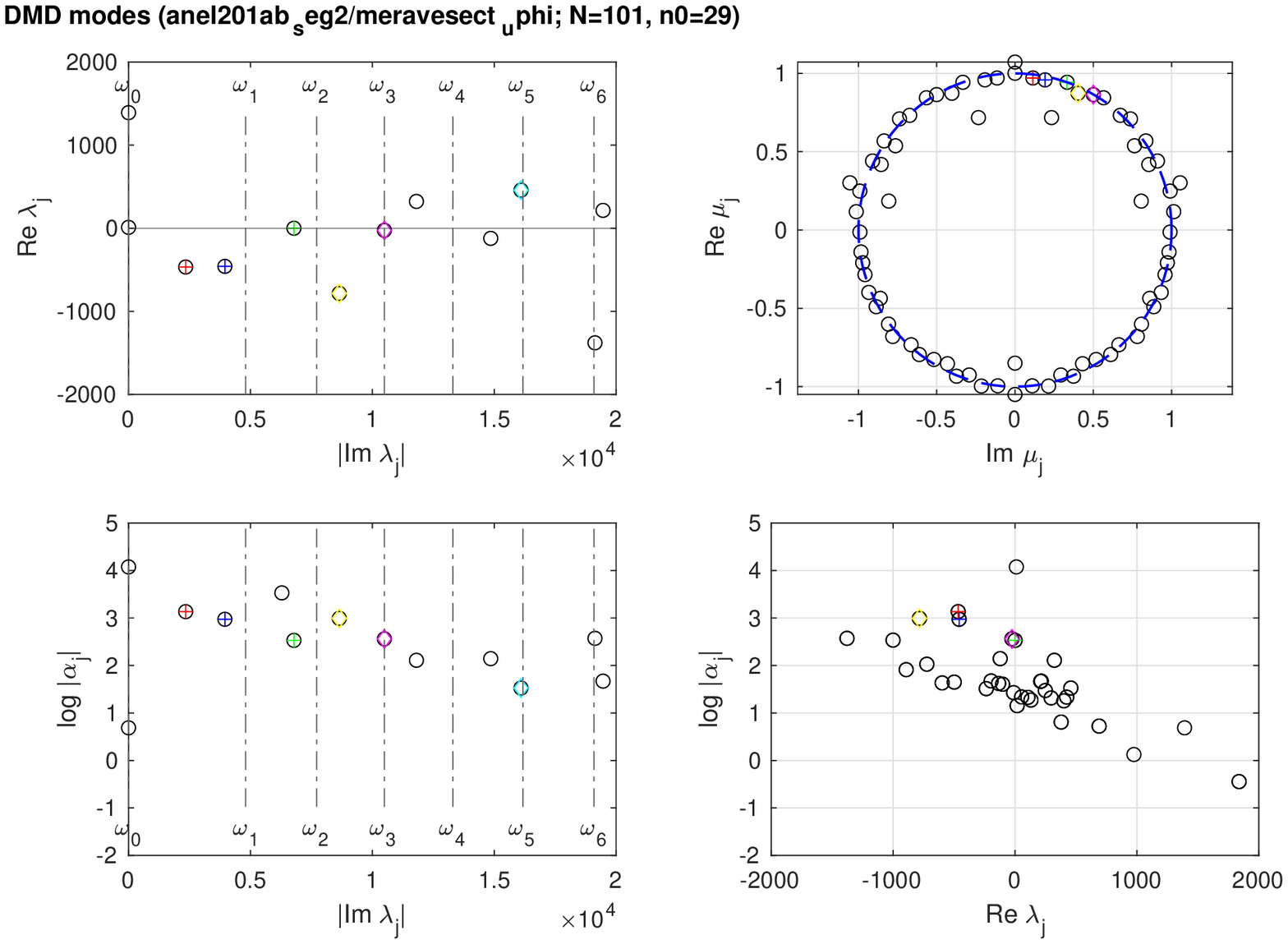} 
\end{center}
\vspace{-2em}
 \caption{DMD eigenvalues (top) and amplitude (bottom) in run E. 
 \rev{Modes 1, 2, 3, 4, 5, and 8 are indicated in red, blue, green, yellow, magenta, and cyan, respectively.}
 The \rev{vertical} dashed-dotted line labelled by $\omega_i$ indicates the frequency of the $i$-th TW normal mode in the metallic region.}
 \label{fig:eigenvalues_jup}
\end{figure}

The analysed time sequence $\langle u_\phi \rangle$, as displayed in figure~\ref{fig:data_jup}, 
 has 87 points in $s$ and \rev{72} snapshots in $t$, giving the rank $r= \rev{72}$ for DMD. 
For residual no greater than \rev{$10^{-7}$}: $\Pi_\tx{loss} <$ \rev{9\% with 60 modes}.
The singular value distribution $\sigma_j$ is now a rather smooth function of $j$ \rev{(not shown)},
 indicating low-order models are relatively ineffective.
The DMD distinguishes distinct modes in the vicinity of zero growth rate (fig.~\ref{fig:eigenvalues_jup}): 
 in the frequency range up to $\omega_6$, 
 there are 10 modes having the quality factor $Q$ higher than 5.
In a similar manner to the magnetoconvection case earlier,
 we rank those DMD modes in their nonzero frequencies;
 `Mode 1' for the slowest mode, `Mode 2' for the secondly slowest mode, and so on.
\rev{Only relevant modes are indicated by colour in the figure.}

\begin{figure}[h]
\begin{center}
 \includegraphics[bb= 25mm 4mm 80mm 83mm, clip, width=0.3\linewidth]{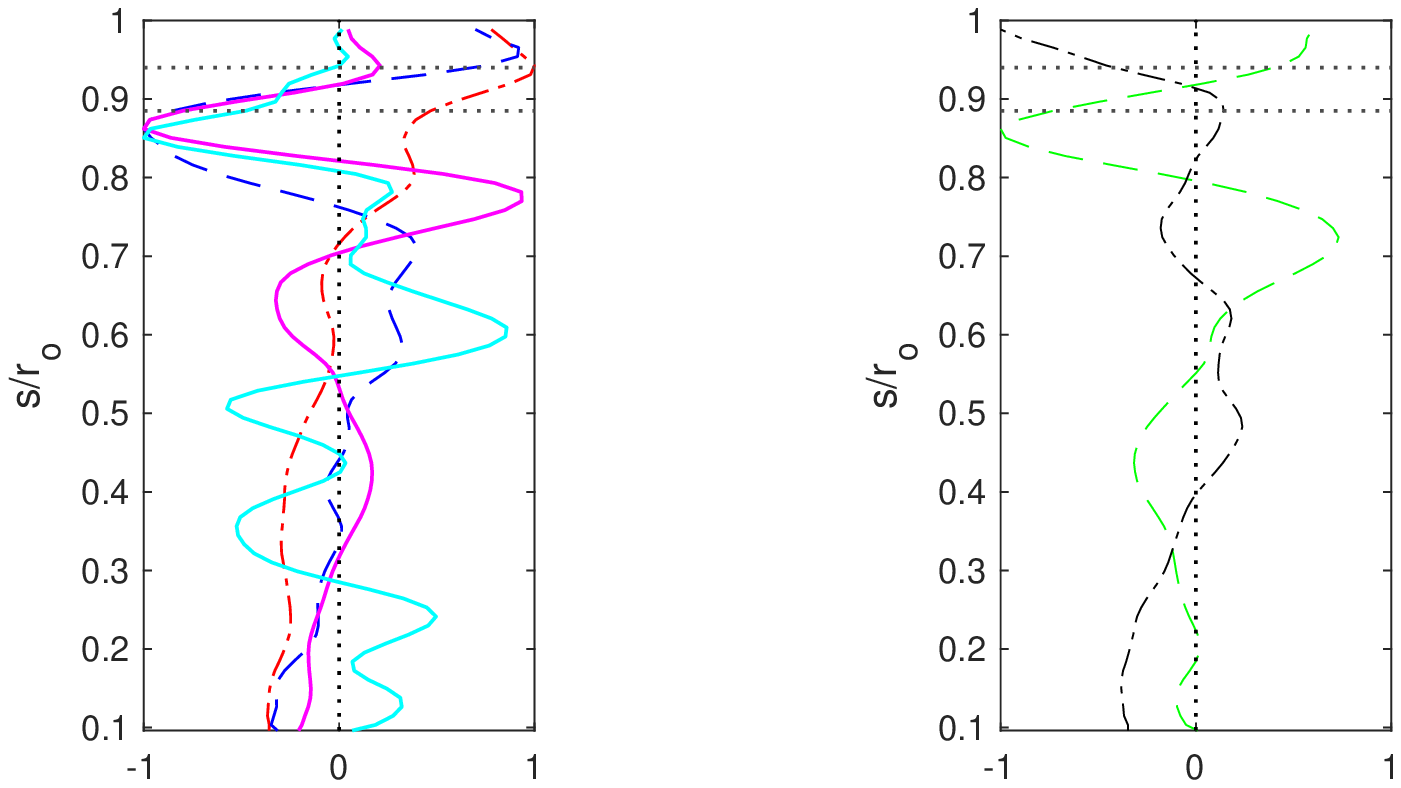}
\end{center}
\vspace{-2em}
 \caption{Profiles of selected DMD eigenfunctions in run E. 
 \rev{The red dashed-dotted, blue dashed, magenta solid, and cyan solid curves represent the ones for Mode 1, 2, 5, and 8, respectively.} 
 The horizontal dotted lines indicate the radii $s/r_\tx{o} =$ 0.885 and 0.94. Only outside the TC is shown.}
 \label{fig:eigenfunctions_jup}
\end{figure}

\begin{figure}[h]
\begin{center}
 \includegraphics[bb= 15mm 12mm 101mm 185mm, clip, width=0.58\linewidth]{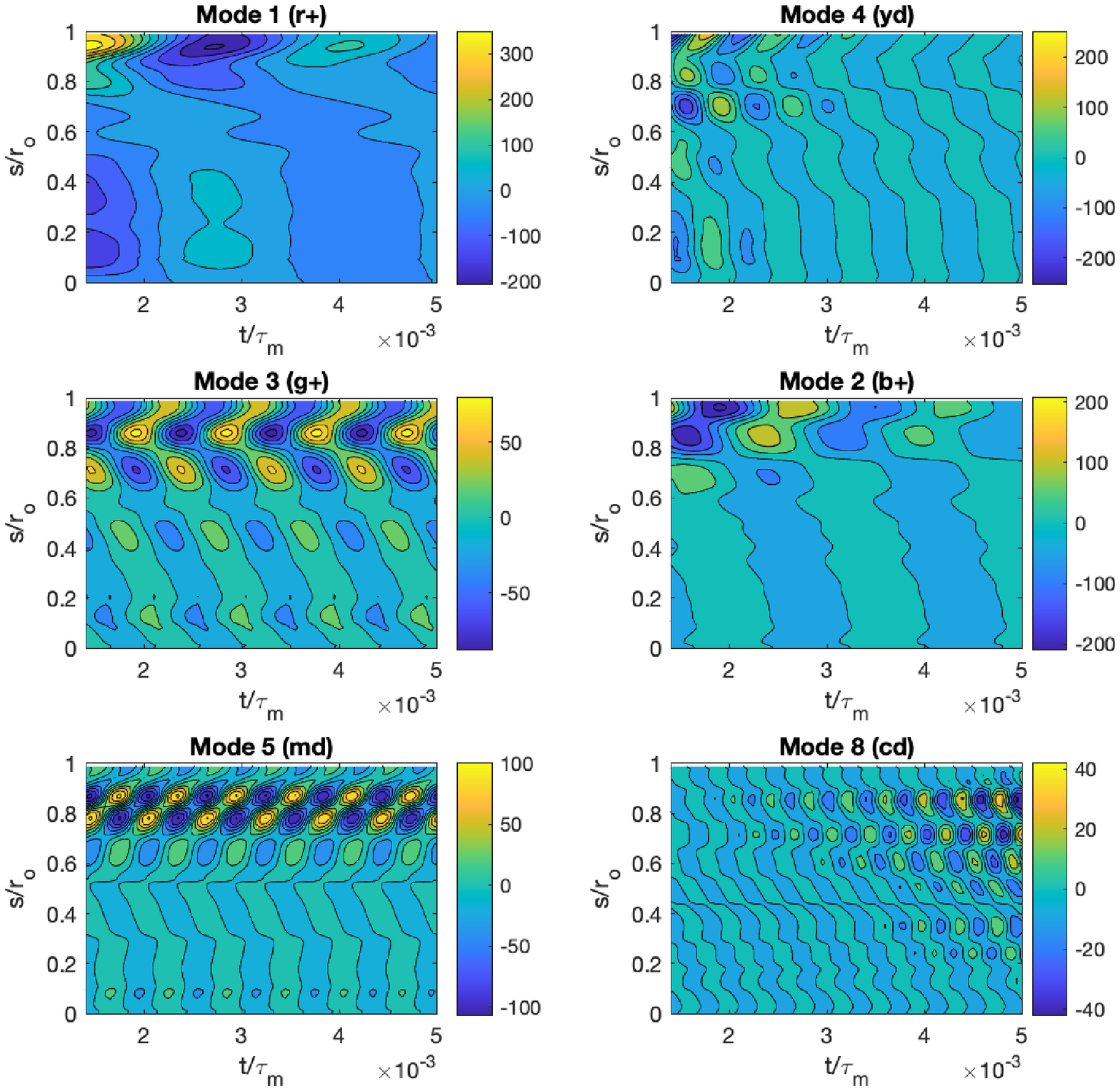}
\end{center}
\vspace{-2em}
 \caption{Spatiotemporal structures of DMD \rev{Modes 1, 3, and 5 (from top to bottom)} in run E.} 
 \vspace{1em}
 \label{fig:eigenfunctions_s-t_jup}
\end{figure}

These modes may be classified as \rev{internal, transition-related, and coupled} modes, 
 depending on the radii at which the DMD eigenfunction has peaks, \rev{as follows.
Their $s$-profiles and spatiotemporal structure are typically demonstrated in figures~\ref{fig:eigenfunctions_jup} and \ref{fig:eigenfunctions_s-t_jup}, respectively.}
\rev{Modes 1 and 4 (as indicated in red and yellow in fig.~\ref{fig:eigenvalues_jup}, respectively)
 have eigenfunctions maximised in the transition or poorly-conducting layer}, 
 i.e. $s/r_\tx{o} \gtrsim 0.9$, while decaying 
\rev{(see the dashed-dotted curve in fig.~\ref{fig:eigenfunctions_jup} and the top panel in fig.~\ref{fig:eigenfunctions_s-t_jup} for Mode 1)}:
 we realise them transition-related modes. 
Modes 2 (blue) and 3 (green) are classed coupled modes: 
 their eigenfunctions of both are largest \rev{at $s/r_\tx{o} \approx 0.86$,
 near the lower bound of the transition zone},
 and notable \rev{at either larger and smaller radii
 (the blue dashed curve in fig.~\ref{fig:eigenfunctions_jup} for Mode 2)}. 
Interestingly those \rev{coupled} modes appear to propagate from the \rev{radius} both towards the deeper interior and towards the poorly conducting layer
 (the \rev{middle} panel in fig.~\ref{fig:eigenfunctions_s-t_jup} \rev{for Mode 3}).  
This could be interpreted as waves being reflected off and partially transmitted through the transition interface:
 the presence of a conductivity jump may give rise to reflections and transmissions of MHD waves \cite{HTJ19}. 
Modes classed as internal include \rev{Modes 5 (magenta) and 8 (cyan)}, 
 for which the eigenfunction is \rev{suppressed when 
 $s/r_\tx{o} \gtrsim 0.94$ but has peaks
 within the metallic region 
 (the magenta and cyan solid curves in fig.~\ref{fig:eigenfunctions_jup}): 
 they reveal} quasi-stationary standing waves, or oscillations 
 (the bottom panel of fig.~\ref{fig:eigenfunctions_s-t_jup} \rev{for Mode 5}).
\rev{We also remark their frequencies almost match $\omega_3$ and $\omega_5$ of the free TW normal modes \rev{(fig.~\ref{fig:eigenvalues_jup})};
 details of the radial profile cannot be met perfectly.
This, again, highlights the need for better understanding of the TW eigenvalue problem in more realistic situations.}

\vspace{1em}

\vspace{1em}

\vspace{1em}

\section{Summary}

We have demonstrated the standard DMD of DNS data for rotating MHD convection and dynamos, 
 in comparison with normal modes computed with the simple, diffusionless TW models.
The methodology provides us with a global, but straightforward, approach
 to characterise MHD wave excited in the deep interiors of planets,
 for which available data is still limited.
We will further extend our exploration to the data accessible at the surface. 

\vspace{1em}
\section*{Acknowledgments}

We acknowledge support from the Japan Society for the Promotion of Science 
 under Grant-in-Aid for Scientific Research (C) No.~20K04106.
This work was partly 
 supported by the Japan Science and Technology/Kobe University under 
 the Program "Initiative for the Implementation of the Diversity Research Environment (Advanced Type)".
We also thank Susanne Horn, Jon Mound and Chris Jones for discussions and comments.

\end{document}